\documentstyle[12pt,nature,epsf]{article}

\date{}
\begin{document}
\newcommand{\be}{\begin{eqnarray}}
\newcommand{\ee}{\end{eqnarray}}
\newcommand{\etal}{{\it{et al.}}}
\newcommand{\smass}{M_{\odot}}
\newcommand{\br}{{\bf r}}
\newcommand{\bV}{{{\bf v}}}

\pagestyle{empty}



%
%
\title{
Three-Body Affairs in the Outer Solar System
}

%
%

\author{Yoko Funato\\
General System Studies,
University of Tokyo,
    Tokyo 153, Japan\\
\medskip
\small{\tt funato@chianti.c.u-tokyo.ac.jp}\\
Junichiro Makino\\
\bigskip
Department of Astronomy, University of Tokyo, Tokyo 113, Japan\\
%
Piet Hut\\
\bigskip
Institute for Advanced Study,
Princeton, NJ 08540, USA\\
%
Eiichiro Kokubo \& Daisuke Kinoshita\\
National Astronomical Observatory, Tokyo 180, Japan}

%
%

\def\jun#1{{\bf[#1---Jun]}}
\def\piet#1{{\bf[#1---Piet]}}
\def\yoko#1{{\bf[#1---Yoko]}}

\maketitle

\begin{abstract}

{\bf
Recent observations\cite{Burnes2002,Veillet2002,Margot2002a} have
revealed an unexpectedly high binary fraction among the
Trans-Neptunian Objects (TNOs) that populate the Kuiper Belt.  The TNO
binaries are strikingly different from asteroid binaries in four
respects\cite{Veillet2002}: their frequency is an order of magnitude
larger, the mass ratio of their components is closer to unity, and
their orbits are wider and highly eccentric.
Two explanations have been proposed for their formation, one assuming
large numbers of massive bodies\cite{Weidenschilling2002}, and
one assuming large numbers of light bodies\cite{Goldreich2002}.  We
argue that both assumptions are unwarranted, and we show how TNO
binaries can be produced from a modest number of intermediate-mass
bodies of the type predicted by the gravitational instability theory
for the formation of planetesimals\cite{GoldreichWard1973}.
We start with a TNO binary population similar to the asteroid binary
population, but subsequently modified by three-body exchange
reactions, a process that is far more efficient in the Kuiper belt,
because of the much smaller tidal perturbations by the Sun.  Our
mechanism can naturally account for all four characteristics that
distinguish TNO binaries from main-belt asteroid binaries.
}
\end{abstract}

\def\simlt{\hbox{ \rlap{\raise 0.425ex\hbox{$<$}}\lower 0.65ex
  \hbox{$\sim$} }}
\def\simgt{\hbox{ \rlap{\raise 0.425ex\hbox{$>$}}\lower 0.65ex
  \hbox{$\sim$} }} 

The TNO binary 1998WW31 has\cite{Veillet2002} a mass ratio
$m_2/m_1 \sim 0.7$, eccentricity $e \sim 0.8$, semimajor axis
$a \sim 2\times 10^4$ km, and inferred radii
$r_1 \sim 1.1r_2 \sim 10^2$ km, hence $a/r_1 > 10^2$, in stark
contrast to main belt asteroid binaries\cite{Margot2002b},
where $m_2/m_1 \ll 1$, $e \sim 0$, and $a/r_1 \simlt 10$.

Asteroid binaries are probably formed by collisions\cite{Merline2003},
as in the leading scenario for the formation of the
Moon\cite{HartmannDavis1975,Kokubo2000}.  The
observed characteristics, $m_2/m_1 \ll 1$, $e \sim 0$, and
$a/r_1 \simlt 10$, are all natural consequences of this
scenario\cite{Durda2001}.  For a different scenario for 1998WW31,
we can look at dynamical binary formation in star clusters, where
there are three channels:
1) tidal capture\cite{FPR1975};
2) three-body binary formation\cite{Heggie1975};
and 3) exchange reactions\cite{Heggie1975}.

Channel 1 is analogous to the standard scenario for asteroid binary
formation.  It will indeed occur: each TNO has grown through
accretion, and much of this accretion has happened through collisions
with an object comparable in mass to the growing TNO
itself\cite{KokuboIda1997, Makino1998}.

Channel 2 would require a near-simultaneous encounter of three massive
objects with low enough velocities to allow an appreciable chance to leave
two of the objects bound.  For this to work, the random velocities of
the most massive objects should be significantly lower than their
Hill velocities.  Under such conditions, this channel could play a
role, as pointed out by Goldreich {\it et al.}\cite{Goldreich2002}, 
who assumed that there are $\sim10^5$ 100 km--sized object embedded
in a sea of small ($<1 {\rm km}$) objects.  This assumption, however,
is at odds with Goldreich and Ward's theory for the formation of
planetesimals\cite{GoldreichWard1973} through gravitational instability,
and it is hard to see how objects in the Kuiper Belt could form from
non-gravitational coagulation, because the time scales are far too
long\cite{Wetherill1990}.  In contrast, the gravitational instability
theory predicts the size of the initial bodies to be $10-100 {\rm km}$.
Starting with these larger bodies would make channel 2 ineffective,
because the velocity dispersion would be higher than the Hill
velocity\cite{KokuboIda1997,KenyonLuu1998}.

Recently, Weidenschilling\cite{Weidenschilling2002} proposed a
variation on the idea of using interactions between three unbound
bodies in order to create a binary.  He studied how a third massive
body could capture the merger remnant from a collision of two massive
bodies if the third body were near enough during the time of the
collision.  This mechanism seems unlikely to work, however, since it
requires a number density of massive objects about two orders of
magnitude higher than the value consistent with present
observations\cite{Goldreich2002}.

Goldreich {\it et al.}\cite{Goldreich2002} have proposed another
mechanism, based on the dynamical friction from a sea of smaller
bodies that can turn a hyperbolic encounter between two massive
bodies into a bound orbit under favorable conditions.  Effectively,
this mechanism makes use of a superposition of three-body encounters,
since each light body interacts independently with the two heavier
ones, and in that sense it is another variant on channel 2.
As we mentioned above, the gravitational instability theory for the
formation of planetesimals\cite{GoldreichWard1973} would exclude the
existence of such a sea of small objects, and since the alternative
theory of nongravitational agglomeration does not seem to work, we
will explore the consequences of dropping channel 2.

Channel 3 can operate on the binaries formed through channel 1, so we
should check whether channel 1 and 3 together produce the right
binaries in the right numbers.

Starting with the first task, consider a relatively massive TNO
primary in a binary orbit with a much less massive secondary.  If the
binary encounters a particle with a mass $m$ that is comparable to the
mass of the primary component ($m_1 \sim m \gg m_2$), the most likely
result is an exchange reaction, in which the incoming object replaces
the original secondary\cite{Spitzer1987}. Figure 1 shows an example of
such a reaction.

\begin{figure}
\begin{center}
\leavevmode
\epsfxsize 10cm
\epsffile{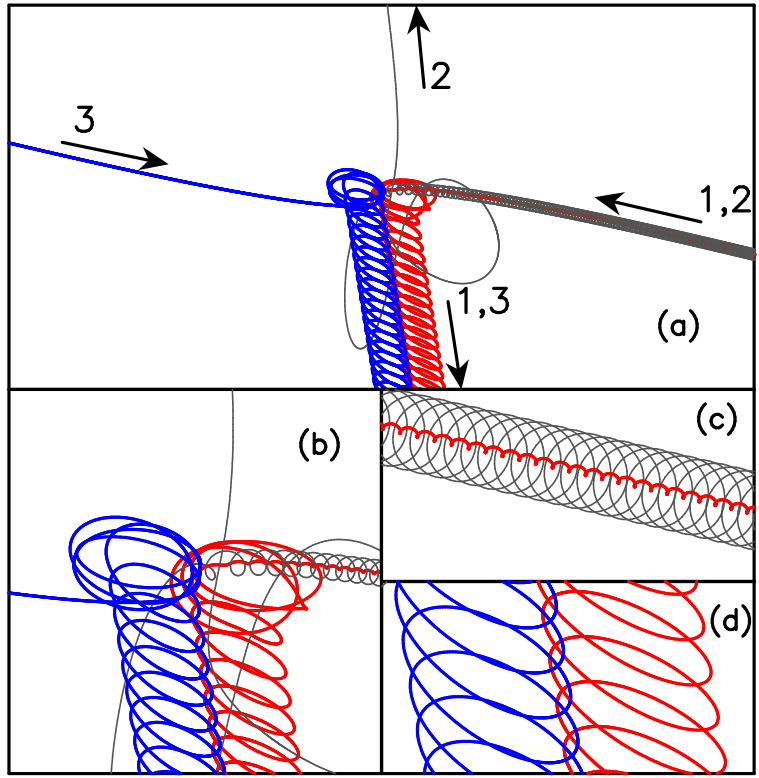}
\end{center}
\label{fig:1a}

\caption{
{\narrower
 An example of a binary--single-body exchange interaction, in the
`(massive, light) meets massive' category discussed in this paper.
Bodies 1 and 2 have masses $m_1=1$ and $m_2=0.1$, respectively,
forming a binary with an initially circular orbit.  Body 3, with mass
$m_3=1$, encounters the binary on an initially parabolic orbit.  In
panel (a), the whole scattering process is shown.  Panel (b) shows the
complex central interaction in more detail, while panels (c) and (d)
show the orbits of the initial and final binary, respectively.
Note that the final binary orbit is highly eccentric and much wider
than the initial circular binary orbit.
}
}
\end{figure}

The binding energy of the binary will not change much during the
exchange, hence $m_1m_2/a_0 \approx m_1m/a$ where $a$ is the new
semimajor axis after the exchange.  This implies
$a/a_0 \approx m/m_2 \gg 1$.  Under the impulse approximation, the
interaction happens in a space small compared to the distance $a_0$ to
the primary.  Conservation of specific angular momentum of the system
gives $m_2a_0(1-e_0) \approx ma(1-e)$ which gives $1-e \approx m_2/m \ll 1$.

We have run a series of scattering experiments to obtain the relevant
cross sections, for an initial binary with mass ratio of $20:1$ and
semimajor axis $a_0=20r_1$, where $r_1$ is the radius of the primary.
These values are typical for main-belt binary asteroids, with
$m_2/m_1 < 0.1$, and separations $5-40$ times the radius of the primary.
We choose parabolic relative orbits for the single body approaching
the binary, with periastron distances uniformly distributed between 0
and $20a_0$.  We only followed the system as long as all three bodies
stayed within their Hill radius, $1000a_0$.

Table 1 gives cross sections for processes in which initial binary
membership is altered.  Channels (a), (c) and (e) result in binaries
with two massive components, and together comprise about 80\% of the
total cross section.  We checked these results through a comparison
with the starlab three-body scattering package\cite{McMH1996}.

\begin{table}
\label{tbl:CrossSection}
\begin{center}
\begin{tabular}{cccccccc}
\hline\hline
channel: & (a) & (b) & (c)  & (d) & (e) & (f) \\
\hline\hline
process: & (1,3),2  & (2,3),1 & (1+2,3) &
 (1+3,2)  & (2+3,1) & no binary \\
\hline
$\sigma v^2$: & 12.1 & 1.3 & 0.9 & 1.3 & 0.9 & 1.2\\
\hline
\end{tabular}
\end{center}
{
Table 1:
Cross sections $\sigma$ for various configuration-changing channels in
binary--single-body scattering.  The gravitational focusing factor
$v^2$ is scaled out in order to obtain finite values in the parabolic
limit, where $v$ is the initial relative velocity between binary and
single body at infinity.  We use units in which $G=m_1=m_3=a=1$,
where $G$ is the gravitational constant, $m_1$ and $m_3$ are the
masses of the heaviest body in the binary and the single body,
respectively, and $a$ is the initial semi-major axis of the binary.
The mass of the lighter body in the binary is $m_2=0.05$.
The radii are $r_1=r_3=0.05$ and $r_2=r_1(m_2/m_1)^{1/3}\approx 0.01842$.
The scattering processes are coded as follows: $(x,y)$ indicates a
binary in the final state with components $x$ and $y$, while a $p+q$
indicates the product of a merger between bodies $p$ and $q$.  A
single body $z$ in the final state is indicated by $(,),z$.
The physical meaning of the six channels is as follows:
(a) an exchange reaction resulting in a massive--massive binary;
(b) an exchange reaction resulting in a massive--light binary;
(c) a merger resulting in a massive--massive binary;
(d) a merger resulting in a twice-as-massive--light binary;
(e) a merger resulting in a massive--massive binary;
(f) no binary is left, after three-body merging or two-body merging
followed by escape.
}

\end{table}

In figure 2 the distribution for the semi-major axis is strongly
peaked at $a=20$, in good agreement with the simple argument presented
above.  Similarly, the eccentricity peaks at 0.95, as expected.  We
assumed $r_1=75 {\rm km}$, the estimated radius of the primary of
1998WW31.  In figure 3, the orbital elements of 1998WW31 are
consistent with the binary having formed through the processes modeled
here.

\begin{figure}
\begin{center}
\leavevmode
\epsfxsize 10cm
\epsffile{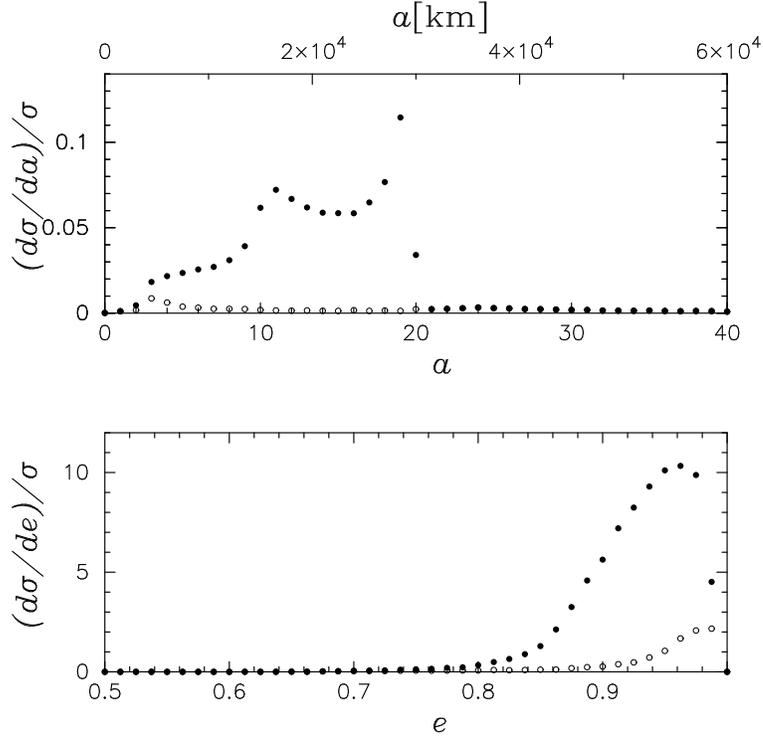}
\end{center}
\label{fig:2}

\caption{ Normalized differential cross sections for the formation of
a `massive-massive' binary, under the conditions specified in the text
(channels a, c and e in table 1), with respect to the semi-major axis
$a$ (top panel), and eccentricity $e$ (bottom panel) of the final
binary.  The initially circular binary has $a=1$ in the dimensionless
units used for $d\sigma/da$, while the physical units are given for
reference at the top of the figure.  The filled points are the total
values for the differential cross sections, while the open circles are
the contributions from the merger channels (c and e in table 1).  Note
the double-peaked structure in the top panel: the sharp peak toward
$a\sim 20$ arises from non-resonant exchanges, where the final binary
has an energy comparable to that of the initial binary; the broad peak
around $a\sim 10$ arises from resonant exchanges, where the memory of
the initial binary is wiped out, leading on average to more strongly
hyperbolic escape in which a harder binary is formed.  }
\end{figure}

\begin{figure}[ht]
\begin{center}
\leavevmode
\epsfxsize 10cm
\epsffile{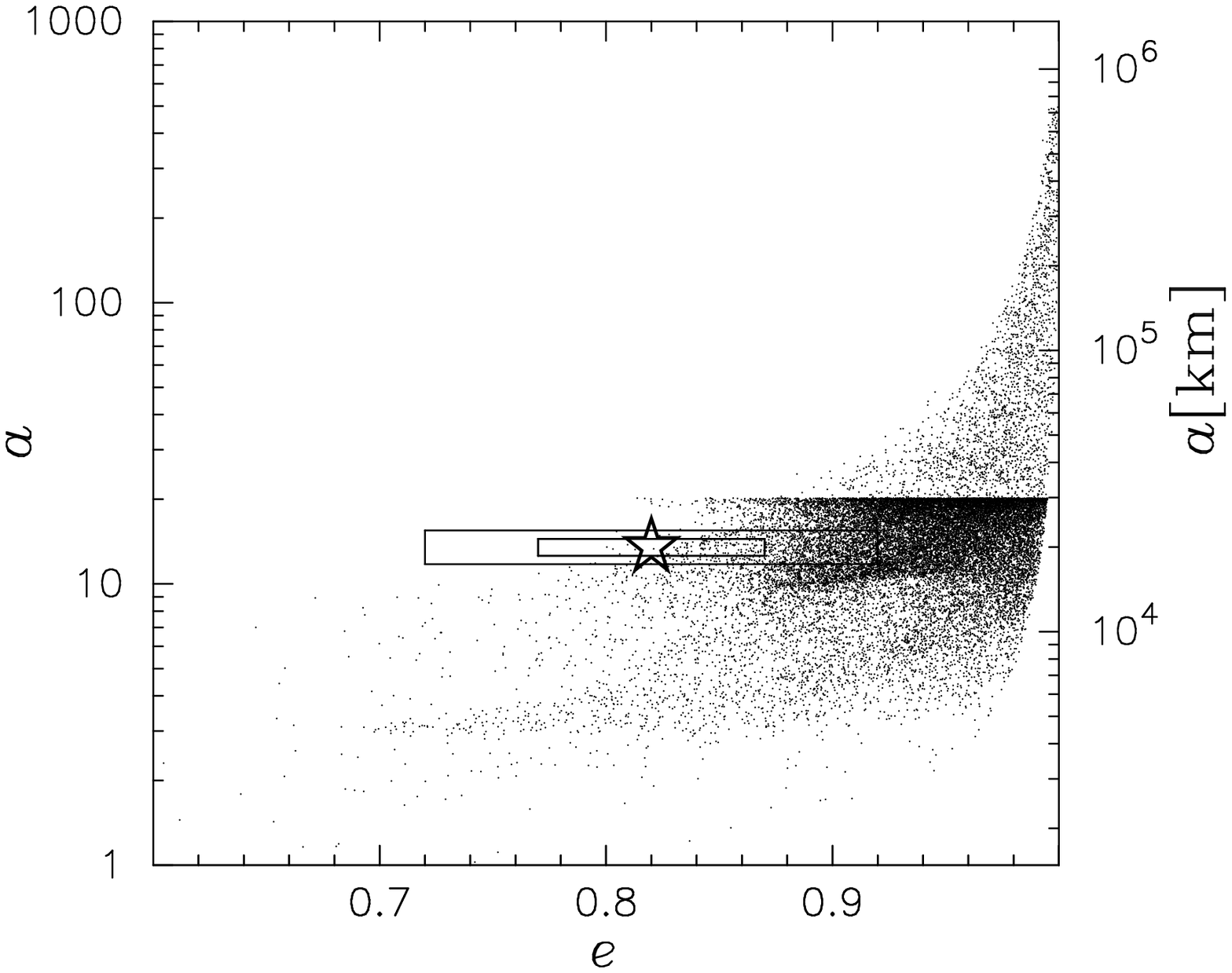}
\end{center}
\caption{Orbital properties of `massive-massive' binaries formed in
our scattering experiments: $a$ and $e$ have the same meaning and
units as in fig. 2.  Contributions from exchange reactions, channel (a)
in table 1, are limited by energy conservation to $a \simlt 20$,
and give rise to the horizontal rim in the middle of the figure.
Contributions involving mergers, channels (c) and (e) in table 1, can
lead to $a$ values all the way to the Hill radius $a \approx 10^3$, but
are limited by angular momentum conservation to increasingly high $e$
for increasing $a$.  The star symbol shows the observed orbit for
1998WW31.  Boxes around the star indicate the observational 1- and
2-$\sigma$ error bars.
}
\end{figure}

We now confront our second task: to check whether
the exchange channel is efficient enough to produce the observed
binaries.
Starting with TNOs of intermediate
mass, as predicted by Goldreich and Ward's theory for the formation of
planetesimals\cite{GoldreichWard1973}, the heaviest TNOs will accrete
mass primarily through collisions with TNOs of comparable
mass\cite{KokuboIda1997, Makino1998}.
Many of these collisions are of the `giant impact'
type that form a tight circular strongly
unequal-mass binary (channel 1).
Let us
estimate what fraction of encounters between comparable-mass
TNOs will give rise to `giant impact' type binaries, and how long such
binaries survive on average before they are destroyed again.

We assume that one in three collisions between comparable TNOs gives
rise to a binary.  When no binary is produced, we have to wait for a
typical time $T$ until another major collision occurs.  When a binary
is formed, gravitational focusing implies a cross section for
three body interactions of order $a_0$.  Therefore, our
newly-formed binary will undergo an exchange reaction on a
time scale $(r/a_0)T \ll T$, leading to a significant increase in $a$.
Strong three-body interactions will subsequently occur on a much
shorter time scale $(r/a)T \ll T$.  As a result, the semimajor
axis will shrink systematically, while the `thermal'
distribution $f(e) = 2e$ favors high eccentricity\cite{Heggie1975}.

When the orbit becomes small enough, $r/a \sim 0.03$, the chance for
collisions in resonant encounters becomes
significant\cite{HutInagaki1985}.  Let us assume that an exchange
reaction turns a `giant impact' binary into a binary with a semi-major
axis of $a \sim 300r$.  Each subsequent strong encounter will on
average decrease $a$ by a factor\cite{HH2003} $\sim 1.2$.  After a
dozen encounters, $a \sim 30r$ and collision is likely to occur.
The time scale for each encounter to occur is $\sim (r/a)T$.  The
waiting time for the last encounter in this series to occur is
$(1/30)T$, while each previous waiting time was less by a factor 1.2.
Summing this series, we get a total waiting time of
$(T/30)/(1 - (1/1.2)) = 0.2T$ before a collision between two or three
massive TNOs.  If all three collide, we are back
where we started, and the resulting system may be a single body (with
an assumed chance of 2/3) or a strongly unequal-mass binary (chance 1/3).
If two of the bodies collide, the third one may remain in orbit, or it
may escape.
In the latter case, we again are back where we started.  In the former
case, we still have an equal-mass and likely highly eccentric wide binary.

Under these assumptions, in 1/3 of the cases, we wind up with an
equal-mass TNO binary with the observed properties for a period $\sim 0.2T$,
compared to a 2/3 chance to wind up with a single TNO for a period
$\sim T$.
This allows us to derive the rate equation for the formation and
destruction of the binaries. 
If we denote by $N_S$ and $N_B$ the number of single bodies and the
number of binaries, respectively, we have
\begin{eqnarray}
\frac{dN_B}{dt} &=& \frac{1}{3}N_S \, -
                    \, \frac{1}{0.2} \,\frac{2}{3}N_B\nonumber\\
\frac{dN_S}{dt} &=& -\,\frac{dN_B}{dt}\nonumber
\end{eqnarray}
if we measure time in unit of $T$. So for the stationary state we have
$dN_B/dt = dN_S/dt=0$, and $N_B=0.2N_S/2=0.1N_S$. Therefore, the
binary fraction is $\sim 10$\%.  When accretion in the Kuiper belt
region diminished, the number of single and binary objects was frozen,
with a ratio similar to this steady-state value.

While our arguments are only approximate, it is clear that after
cessation of the accretion stage at least several percent or more of
the TNOs were accidentally left in such a binary phase.  The fact that
more than 1\% of the known TNOs are found to be in wide roughly
equal-mass binaries is thus a natural consequence of {\it any}
accretion model {\it independent of the assumed parameters} for the
density and velocity dispersion of the protoplanetary disk or the
duration of the accretion phase.  As a corollary,
we predict that future discoveries of TNO binaries will similarly show
roughly equal masses, large separations, and high eccentricities.

We conclude that we have found a robust and in fact unavoidable way to
produce the type of TNO binaries that have been found, as long as we
start from the plausible assumption that TNOs were formed through
gravitational instabilities\cite{GoldreichWard1973}.

{\bf Acknowledgements}
We acknowledge helpful comments on our manuscript by Peter Goldreich
and Roman Rafikov.

\newpage


\begin{thebibliography}{99}

\bibitem{Burnes2002}  Burnes, J. A., Science, 297, 942-943 (2002).

\bibitem{Veillet2002} Veillet C., et al., Nature, 416, 711-713 (2002).

\bibitem{Margot2002a} Margot, J-L., Nature, 416, 694-695, (2002).

\bibitem{Weidenschilling2002} Weidenschilling, S. J., Icarus, 160, 212-215 (2002).

\bibitem{Goldreich2002} Goldreich, P., Lithwick, Y., and Sari, R.,
Nature, 420, 643-646 (2002).

\bibitem{GoldreichWard1973} Goldreich, P. and Ward, W. R., ApJ, 183,
1051 (1973).

\bibitem{Margot2002b} Margot, J-L., Nolan, M. C., Benner, L. A. M.,
Ostro, S. J., Jurgens, R. F., Giorgini, J. D., Stade, M. A., and
Campbell, D. B., Science, 296, 1445-1447, (2002).

\bibitem{Merline2003} Merline, W. J. et al. in Asteroids III (eds Bottke, W. F., Cellino, A.,
   Paolicchi, P. \& Binzel, R. P.) (in the press).

\bibitem{HartmannDavis1975} Hartmann, W. K. and  Davis,D. R., Icarus,
24, 504-515 (1975).

\bibitem{Kokubo2000} Kokubo, E., Ida, S., Makino, J. Icarus, 148, 419--436,  (2000).

\bibitem{Durda2001} Durda, D. D., Bottke, W. F., Asphaug, E., and
Richardson, D. C., Bull. Am. Astron. Soc., 33, 1134 (2001).
   
\bibitem{FPR1975} Fabian, A.C., Pringle, J.E. \& Rees, M.J.,
MNRAS 172, 15P (1975).

\bibitem{Heggie1975} Heggie, D.C., MNRAS, 173, 729 (1975).

\bibitem{KokuboIda1997} Kokubo, E. and Ida, S., Icarus, 131, 171 (1998).

\bibitem{Makino1998} Makino, J., {Fukushige}, T., {Funato}, Y., \&
{Kokubo}, E., {New Astronomy}, 3, {411-416} (1998)

\bibitem{Wetherill1990} Wetherill, G.W., Annu. Rev. Earth Planet. Sci. 18,
205 (1990).

\bibitem{KenyonLuu1998} Kenyon, S. J.and Luu, J. X., Astronomical
Journal, 115, 2125 (1998).

\bibitem{Spitzer1987}Spitzer, L., ``{Dynamical evolution of globular
clusters}'', {Princeton, NJ, Princeton University Press}, (1987).

\bibitem{McMH1996} McMillan, S. \& Hut, P., Astrophysical
Journal, 467, 348-358  (1996).
   
\bibitem{HutInagaki1985} Hut, P. and {Inagaki}, S., Astrophysical
Journal, 298, 502-520  (1985).

\bibitem{HH2003} Heggie, D.C. \& Hut, P. {\it The Gravitational
Million-Body Problem} [Cambridge Univ. Pr.], Ch. 23 (2003).

\end{thebibliography}
\end{document}